# TESTING DIFFERENT LOG BASES FOR VECTOR MODEL WEIGHTING TECHNIQUE

Kamel Assaf

Programmer, Edmonton, AB, Canada

## ABSTRACT

*Information retrieval systems retrieves relevant documents based on a query submitted by the user. The documents are initially indexed and the words in the documents are assigned weights using a weighting technique called TFIDF which is the product of Term Frequency (TF) and Inverse Document Frequency (IDF). TF represents the number of occurrences of a term in a document. IDF measures whether the term is common or rare across all documents. It is computed by dividing the total number of documents in the system by the number of documents containing the term and then computing the logarithm of the quotient. By default, we use base 10 to calculate the logarithm. In this paper, we are going to test this weighting technique by using a range of log bases from 0.1 to 100.0 to calculate the IDF. Testing different log bases for vector model weighting technique is to highlight the importance of understanding the performance of the system at different weighting values. We use the documents of MED, CRAN, NPL, LISA, and CISI test collections that scientists assembled explicitly for experiments in data information retrieval systems.*

## 1. INTRODUCTION

Just not from long time ago libraries were the right place to search for an information. People usually ask the librarian for a particular book, content or search among the available books. Consequently, this becomes a tedious and inefficient task to find what you really need. With the advancements of technology and computing, it is now possible to create a digital information retrieval system capable of letting users input queries, process it and deliver relevant information as output to the user[1, 13]. It is used to find relevant information to the user query [9, 13]. Relevant information is any set of items such as images, documents, videos and other types of data. For the satisfaction of end users, it is expected by the system to retrieve precise and accurate information. But, it is not always that the system returns with the desired data. Hence, allowing the expansion of research in the subject. The main goal of this research is to elevate the performance of information retrieval systems by finding new values as we calculate the TFIDF weighting model. Modifying the default log base when calculating the IDF portion of the equation could potentially make the system retrieve more precise and accurate information.

Many models exist for information retrieval systems such as the vector model[4, 13] and the probabilistic model[12, 13]. In particular to this research, we will explore the vector model and its common method of weighting called TFIDF which is the product of Term Frequency and Inverse Document Frequency[4, 8]. Term Frequency is the representation of a word occurrences in a text and the Inverse Document Frequency define if a word is common or rare across all documents. This research looks into a different way to improve the information retrieval system results. My idea is to change the log base when calculating the IDF from the default base 10 to test a range of variant log bases ranging from 0.1 to 100.0. To solve this problem, we explore a mathematical method used to calculate the log of a value using a different base.

## 2. IR SYSTEM REVIEW

This section is reserved to explore some of the components of an information retrieval system and its characteristics. I did develop my own system in order to made this research possible and understand the subject in deep.

### 2.1. Test Collection

Many standard collections are available for testing information retrieval system[2]. Table 1, shows the five test collections that was used to evaluate this research, it contains information regarding the size, number of documents and the number of terms in the test collection. Each one of the test collections comes with a queries list and a relevance judgment list that indicates the relevant documents for each query.

Table 1. Small Test Collections: Size, Documents and Terms

| Test Collection | Size in Megabytes | No of Documents | No of Terms |
|---|---|---|---|
| MED | 1.05 | 1,033 | 8,915 |
| CRAN | 1.40 | 1,400 | 4,217 |
| NPL | 3.02 | 11,429 | 7,934 |
| LISA | 3.04 | 6,003 | 11,291 |
| CISI | 1.98 | 1,460 | 5,591 |

### 2.2. Indexing

Indexing refers to the steps of eliminating stopwords, removing connectives, punctuations and stemming words by removing its affixes and suffixes[11].

#### 2.2.1. Stopwords

In the first step, we read all the documents and compare its words with a list of stopwords found in a stoplist[1]. The stopwords list consists of adjectives, adverbs, and connectives such as *another*, *again*, *between*, *mostly*, *the*, *a*, *in* and others are considered unnecessary words, because they have different functionality in the grammar, they are mostly used to connect words and create sentences. This step also include the removal of punctuations and special characters from the documents.

#### 2.2.2. Stemming

The stemming process is the technique used to transform particular words by removing the suffixes from the word[7]. A well-known technique is the Porter stemming algorithm, developed at the University of Cambridge[11]. A simple example is the word ***playing*** that becomes ***play*** and ***books*** becomes ***book***, removing the endings ***-ing*** and ***-s***respectivelly. Another example below shows the removal of some suffixes from the words ending:

SSES → SS

- care**sses** = caress
- pon**ies** = poni
- **ties** = ti

IES → I

- pon**ies** = poni

ES → *in this case, remove the ending "ES"and add nothing to it.*

- beach**es** = beach
- bush**es** = bush

The stemming algorithm cleans and make words smaller in size and it is proven in previous research that using it will improve the performance of your system[1, 7].

## 2.3. Vector Model

This model represents documents as vectors in space[2]. In this model, words are referred as terms and every indexed term extracted from a document becomes an independent dimension in a dimensional space. Therefore, any text document is represented by a vector in the dimension space[4]. Various techniques for weighting exists[4]. One of the primary weights is the TFIDF weighting measure. It is composed of a local and global weight. For instance, Term Frequency (TF) is the number of a term occurrences in a document. The global weight is the document frequency (DF), it defines in how many documents the word occurs. Inverse document frequency (IDF) has DF scaled to the total number of documents in the collection as shown the equation below [12, 13]:

$$IDF_t = log_b \left(\frac{N}{DroFreq(t)}\right) \qquad (1)$$

Where *N* is the total number of documents in the test collection, *DocFreq(t)* is the total number of documents containing term *t* and *b* is calculated using base ten. Finally, TFIDF is computed as shown in below equation [12, 13]:

$$w_t = tf_{i,j} * idf_i \qquad (2)$$

This technique of weighting is applied for both the terms in the test collection and also to the terms in the query.

## 2.4. Querying

The set of documents in a collection are a set of vectors in space[1]. To compare a query submitted by the user to the documents, we use the cosine similarity[3, 13] as shown below:

$$\cos\theta = \frac{\sum_{i=1}^{n} a_i \times b_i}{\sqrt{\sum_{i=1}^{n} a_i^2} \times \sqrt{\sum_{i=1}^{n} b_i^2}} \qquad (3)$$

In the above equation, *a* and *b* represent the words in the query and in the document respectivelly. It was used to apply the cosine rule for each of the documents against the submitted query. The results indicate the top-scoring documents for a query [1]. Documents with the highest similarity value are considered the most relevant. After applying this rule for each document in the test collection, we use the above equation to calculate the query and rank the relevant documents related to the query in the space vector, a vector space relies between 0.0 and 1.0.

## 2.5. Assessing

For a better understanding of the research and the results, it was used some sort of assessment that clarifies the results. In the case of information retrieval, the process involves the use of two files that comes with the test collection. Usually, a file which contains a list of predefined queries[5] and another file containing a judgment list that has a query number that indicates the number of the query in the queries list file and a list of relevant documents for that specific query. Both files has been assembled by researchers to make further assessments possible in information retrieval system evaluations[3]. When a researcher submits a query to the information retrieval system, a list of documents are usually returned, for each iteration in the execution of a query in the queries list, the results are then used to calculate precision and recall values[6]. The below equations [12] shows how precision is calculated:

$$precision = \frac{relevant\ items\ retrieved}{retrieved\ items} \qquad (4)$$

On the other hand, we also calculate the recall which gives a view on how effective is your system.

$$recall = \frac{relevant\ items\ retrieved}{relevant\ items} \qquad (5)$$

Calculating the precision and recall values could be done in parallel as they don't rely on values of each other at the time of their calculations. The below figure, shows two possible rankings together with the recall and precision values calculated at every rank position for a query that has six relevant documents.

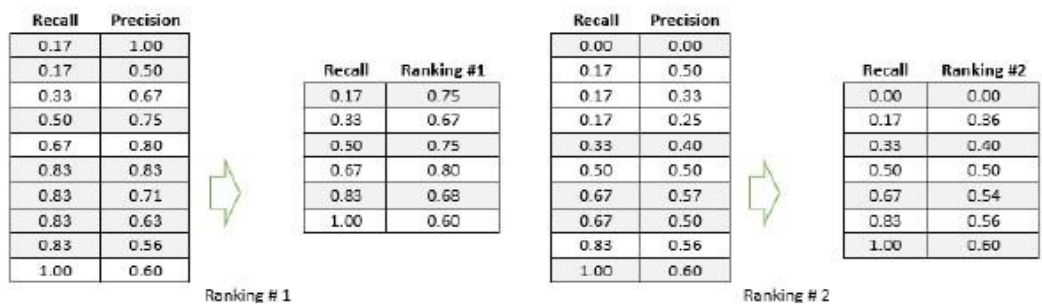

Figure 1. Precisions values averaged for Ranking #1 and Ranking #2

Consequently, precision and recall values are plotted to give a precision-recall curve as shown below:

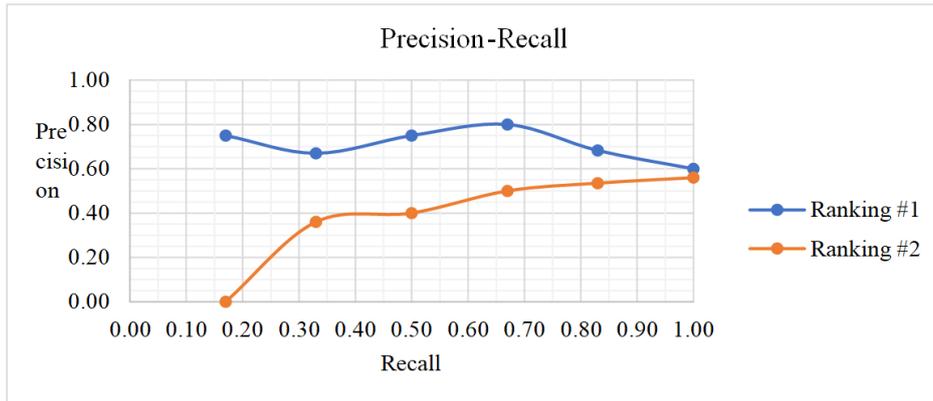

Figure 2. Precision-Recall curve for Rankings #1 and #2

The above example is about ten documents only. However, when we have too many documents, the most typical representation is to set the recall levels to fixed values 0.0, 0.1, 0.2, 0.3, 0.4, 0.5, 0.6, 0.7, 0.8, 0.9, 1.0, then calculate the precision values to the closest fixed recall. Below is an example of assessing many documents:

Table 2. Precision values for many documents at different recall levels

| Recall range 0.00 - 0.05 for 0.0 | | Recall range 0.05 - 0.15 for 0.1 | | Recall range 0.15 - 0.25 for 0.2 | Recall range 0.25 - 0.35 for 0.3 | Recall range 0.35 - 0.45 for 0.4 | Recall range 0.45 - 0.55 for 0.5 | | Recall range 0.55 - 0.65 for 0.6 | Recall range 0.65 - 0.75 for 0.7 | Recall range 0.75 - 0.85 for 0.8 | Recall range 0.85 - 0.95 for 0.9 | | | Recall range 0.95 - 1.00 for 1.0 |
|---|---|---|---|---|---|---|---|---|---|---|---|---|---|---|---|
| 0.00 | 0.00 | 0.02 | 0.11 | 0.13 | 0.20 | 0.39 | 0.42 | 0.52 | 0.54 | 0.64 | 0.70 | 0.81 | 0.86 | 0.89 | 0.90 | 0.96 |
| 1.00 | 0.86 | 0.74 | 0.69 | 0.66 | 0.57 | 0.52 | 0.50 | 0.49 | 0.35 | 0.35 | 0.34 | 0.33 | 0.33 | 0.32 | 0.29 | 0.00 |

It is consider in the above table that all precision values in the recall range 0.00 and 0.05 belong to the recall level 0.0. Likewise, we consider that all precision values that are in the recall range 0.05 and 0.15 belong to the recall level 0.1. Ongoing in the same sequence, all precision values that are in the recall range 0.15 and 0.25 belong to the recall level 0.2 and so on. Table 2 shows how we group the precision values based on the recall ranges and then we average them as shown in table 3.

Table 3. Average Precisions values for all fixed recall levels

| 0.0 | 0.1 | 0.2 | 0.3 | 0.4 | 0.5 | 0.6 | 0.7 | 0.8 | 0.9 | 1.0 |
|---|---|---|---|---|---|---|---|---|---|---|
| 0.867 | 0.675 | 0.570 | 0.520 | 0.500 | 0.420 | 0.350 | 0.340 | 0.330 | 0.313 | 0.000 |

Mean Average Precision (MAP) is a further step to assess the results. MAP is the average of precision values at the fixed recall levels together[1]. The MAP value for the precision values in table 3 is 0.867 + 0.675 + 0.570 + 0.520 + 0.500 + 0.420 + 0.350 + 0.340 + 0.330 + 0.313 + 0.000 = 0.444. MAP@30 is another measure that is more significant than MAP since it indicates how well the system is retrieving relevant documents on the first few pages. MAP@30 is the average of precision values at recall levels 0.0, 0.1, 0.2, and 0.3. MAP@30 for the precision values in table 3 is 0.867 + 0.675 + 0.570 + 0.520 = 0.658.

## 3. Research Methodology

In this paper, a research targeting information retrieval systems has been applied to study opportunities that could lead to further improvements in existing methods for information retrieval systems. The goal is to find new log base values that potentially elevates the performance and effectiveness of the system, which means more accurate and precise results retrieved. Based on books and other researchers, it is used to strongly point that the vector model and the probabilistic model are the most interesting ones and they rely heavily on complex equations to calculate weight values of words in documents[1, 2]. In particular to this paper, the research was focused on the vector model. It was used to change the log base value to calculate the IDF portion of the TFIDF using different log bases varying from 0.1 to 100.0, then compare the results with the default log base 10.

In mathematics, common logarithms calculations are done using the value of 10 as the default base. As we need to change that base for this research, it was used another way to calculate the log, the below equation shows a method that already exists in mathematics that is used to calculate the log using different bases:

$$log_b(x) = \frac{log(x)}{log(b)} \qquad (6)$$

As described in the above equation, we note that it divides the log value of $x$ by the log of the new desired base $b$. Many properties of logarithms exist, but in this case, we focus on the property of changing the base of the logarithm. It was used to implement an algorithm that will allow us to calculate different log bases, simply by diving the log base of the value of idf by the log of the desired base, the algorithm below describes:

1. for $\exists\ base\ (x)\ in\ range\ (0.1 \rightarrow 100.0)$
2. $tf$   = occurrences in document
3. $v$   = docfreq(t)
4. $idf = \log(v) / \log(x)$
5. $tfidf$  = tf * idf

Therefore, for each log base, a new test for the system is considered, using the equation number 6 of this section for idf calculation. This has been applied to all five test collections, in a total of a thousand execution per test collection.

## 4. Research Results

The experiments were applied to the public available small test collections such as MED, CRAN, NPL, LISA, and CISI. I have obtained them from https://ir.dcs.gla.ac.uk/resources/test_collections. In this section we explain the results in a tabular and graph form. Each of below points describes the results found in the five test collections, we classify the results as the best top for MAP and MAP@30, finalizing it with a graph comparison between the best log found, the standard log base 10 and the worst log base found for the test collection.

## 4.1. MED Test Collection

For the MED test collection, it was found significant improvements using log base 0.1, better precision values found at recall levels 0.6, 0.7, 0.8, 0.9 and 1.0 in comparison with the standard log base 10.

Table 4. MED with top 5 log values for MAP

| LOG | \multicolumn{11}{c}{RECALL} | MAP |
|---|---|---|---|---|---|---|---|---|---|---|---|---|
|  | 0 | 0.1 | 0.2 | 0.3 | 0.4 | 0.5 | 0.6 | 0.7 | 0.8 | 0.9 | 1 |  |
| 0.1 | 1 | 0.83 | 0.75 | 0.68 | 0.63 | 0.62 | 0.62 | 0.58 | 0.5 | 0.43 | 0.42 | 0.641818 |
| 0.2 | 1 | 0.83 | 0.75 | 0.68 | 0.66 | 0.63 | 0.62 | 0.5 | 0.47 | 0.43 | 0.42 | 0.635455 |
| 0.3 | 1 | 0.83 | 0.75 | 0.68 | 0.66 | 0.63 | 0.62 | 0.5 | 0.47 | 0.43 | 0.42 | 0.635455 |
| 1.5 | 1 | 0.83 | 0.75 | 0.68 | 0.63 | 0.62 | 0.61 | 0.5 | 0.43 | 0.42 | 0.41 | 0.625455 |
| 2 | 1 | 0.83 | 0.75 | 0.68 | 0.63 | 0.62 | 0.61 | 0.5 | 0.43 | 0.42 | 0.41 | 0.625455 |

Table 5. MED with top 5 log values for MAP@30

| LOG | 0 | 0.1 | 0.2 | 0.3 | 0.4 | 0.5 | 0.6 | 0.7 | 0.8 | 0.9 | 1 | MAP@30 |
|---|---|---|---|---|---|---|---|---|---|---|---|---|
| 0.1 | 1 | 0.83 | 0.75 | 0.68 | 0.63 | 0.62 | 0.62 | 0.58 | 0.5 | 0.43 | 0.42 | 0.815 |
| 0.2 | 1 | 0.83 | 0.75 | 0.68 | 0.66 | 0.63 | 0.62 | 0.5 | 0.47 | 0.43 | 0.42 | 0.815 |
| 0.3 | 1 | 0.83 | 0.75 | 0.68 | 0.66 | 0.63 | 0.62 | 0.5 | 0.47 | 0.43 | 0.42 | 0.815 |
| 1.5 | 1 | 0.83 | 0.75 | 0.68 | 0.63 | 0.62 | 0.61 | 0.5 | 0.43 | 0.42 | 0.41 | 0.815 |
| 2 | 1 | 0.83 | 0.75 | 0.68 | 0.63 | 0.62 | 0.61 | 0.5 | 0.43 | 0.42 | 0.41 | 0.815 |

In the above tables, it was indicated the top five log values, for MAP and MAP@30, it was found that the best MAP is at log 0.1 with a MAP score 0.641, the other two best log values are the 0.2 and 0.3 respectively with both of them having a MAP value of 0.635.

Table 6. MED Compare the best, standard and worst for MAP@30

| LOG | 0 | 0.1 | 0.2 | 0.3 | 0.4 | 0.5 | 0.6 | 0.7 | 0.8 | 0.9 | 1 | MAP@30 |
|---|---|---|---|---|---|---|---|---|---|---|---|---|
| 0.1 | 1 | 0.83 | 0.75 | 0.68 | 0.63 | 0.62 | 0.62 | 0.58 | 0.5 | 0.43 | 0.42 | 0.815 |
| 10 | 1 | 0.83 | 0.75 | 0.68 | 0.63 | 0.62 | 0.61 | 0.5 | 0.43 | 0.42 | 0.36 | 0.815 |
| 0.5 | 1 | 0.83 | 0.75 | 0.68 | 0.63 | 0.62 | 0.58 | 0.5 | 0.43 | 0.42 | 0.36 | 0.815 |

In the above table, we define the values of the best log results, for the base 10 and other bases. A graph then is plotted in below figure with the values from the above table for a better visualization of the results. We can see that log base 0.1 performs better than other bases and compared to base 10 on many recall levels such as 0.6, 0.7, 0.8, 0.9 and 1.0.

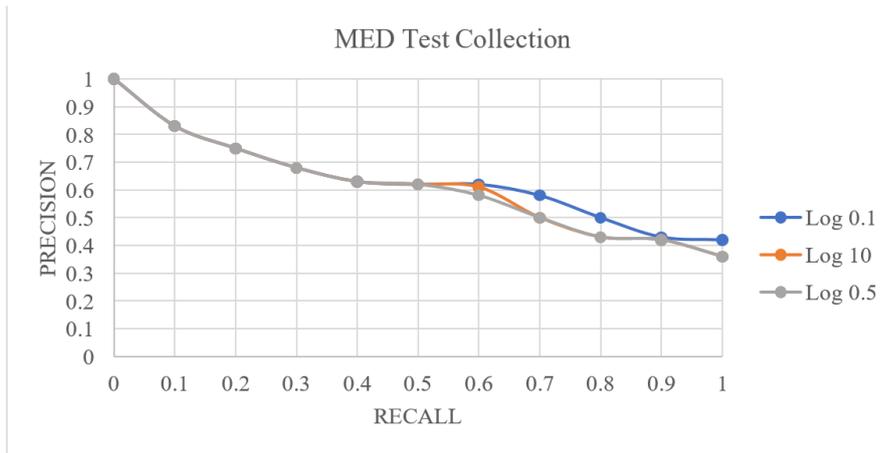

Figure 3. MED Best, Worst and Standard Precision values.

We have determined the logarithm's base, which gives the worst and best performance. When we assess the MED test collection, we find that the system has a MAP@30 equal to 0.815 at all logarithm's bases. The system is at its best performance with a MAP value equal to 0.641 when we apply the TFIDF weighting technique using logarithm at base 0.1. The results are better than the standard logarithm at base 10. It was found the worst performance using log base 0.5.

### 4.2. CRAN Test Collection

In the CRAN test collection, it was found some improvements using log base, higher precision values found at recall levels 0.1, 0.7 and 0.8 in comparison with the standard log base 10.

Table 7. CRAN top 5 log values for MAP

| | RECALL | | | | | | | | | | | |
|---|---|---|---|---|---|---|---|---|---|---|---|---|
| **LOG** | **0** | **0.1** | **0.2** | **0.3** | **0.4** | **0.5** | **0.6** | **0.7** | **0.8** | **0.9** | **1** | **MAP** |
| 0.3 | 1 | 0.88 | 0.83 | 0.8 | 0.78 | 0.76 | 0.75 | 0.73 | 0.7 | 0.68 | 0.68 | 0.780909 |
| 0.2 | 1 | 0.9 | 0.83 | 0.8 | 0.78 | 0.75 | 0.73 | 0.72 | 0.7 | 0.66 | 0.68 | 0.777273 |
| 0.5 | 1 | 0.88 | 0.83 | 0.8 | 0.78 | 0.75 | 0.73 | 0.72 | 0.7 | 0.68 | 0.68 | 0.777273 |
| 1.6 | 1 | 0.88 | 0.83 | 0.8 | 0.78 | 0.75 | 0.73 | 0.72 | 0.7 | 0.68 | 0.68 | 0.777273 |
| 18.7 | 1 | 0.88 | 0.83 | 0.8 | 0.78 | 0.75 | 0.73 | 0.72 | 0.7 | 0.68 | 0.68 | 0.777273 |

Table 8. CRAN top 5 log values for MAP@30

| | RECALL | | | | | | | | | | | |
|---|---|---|---|---|---|---|---|---|---|---|---|---|
| **LOG** | **0** | **0.1** | **0.2** | **0.3** | **0.4** | **0.5** | **0.6** | **0.7** | **0.8** | **0.9** | **1** | **MAP@30** |
| 0.2 | 1 | 0.9 | 0.83 | 0.8 | 0.78 | 0.75 | 0.73 | 0.72 | 0.7 | 0.68 | 0.66 | 0.8825 |
| 0.3 | 1 | 0.88 | 0.83 | 0.8 | 0.78 | 0.76 | 0.75 | 0.73 | 0.7 | 0.68 | 0.68 | 0.8775 |
| 0.5 | 1 | 0.88 | 0.83 | 0.8 | 0.78 | 0.75 | 0.73 | 0.72 | 0.7 | 0.68 | 0.68 | 0.8775 |
| 1.6 | 1 | 0.88 | 0.83 | 0.8 | 0.78 | 0.75 | 0.73 | 0.72 | 0.7 | 0.68 | 0.68 | 0.8775 |
| 18.7 | 1 | 0.88 | 0.83 | 0.8 | 0.78 | 0.75 | 0.73 | 0.72 | 0.7 | 0.68 | 0.68 | 0.8775 |

In above tables, it indicates the top five log values for MAP and MAP@30. It was found the best MAP at log 0.3 with a MAP score 0.7809, the other two best log values are the 0.2 and 0.5 with both of them having a MAP value of 0.7772. For MAP@30 log base 0.2 performs better with a score of 0.8825.

Table 9. CRAN compare the best, standard and worst for MAP@30

| LOG | RECALL | | | | | | | | | | | MAP@30 |
|---|---|---|---|---|---|---|---|---|---|---|---|---|
| | 0 | 0.1 | 0.2 | 0.3 | 0.4 | 0.5 | 0.6 | 0.7 | 0.8 | 0.9 | 1 | |
| 0.2 | 1 | 0.9 | 0.83 | 0.8 | 0.78 | 0.75 | 0.73 | 0.72 | 0.7 | 0.68 | 0.66 | 0.8825 |
| 10 | 1 | 0.88 | 0.83 | 0.8 | 0.78 | 0.75 | 0.73 | 0.7 | 0.68 | 0.66 | 0.68 | 0.8775 |
| 21.7 | 1 | 0.83 | 0.8 | 0.78 | 0.76 | 0.75 | 0.73 | 0.7 | 0.68 | 0.68 | 0.66 | 0.8525 |

We define the values of the best log results compared with the base 10 and others. In the below figure, a graph is plotted with the values from the above table for a better visualization of the results. It indicates that log base 0.2 is really better than at 0.1, 0.8 and 0.9 recall levels.

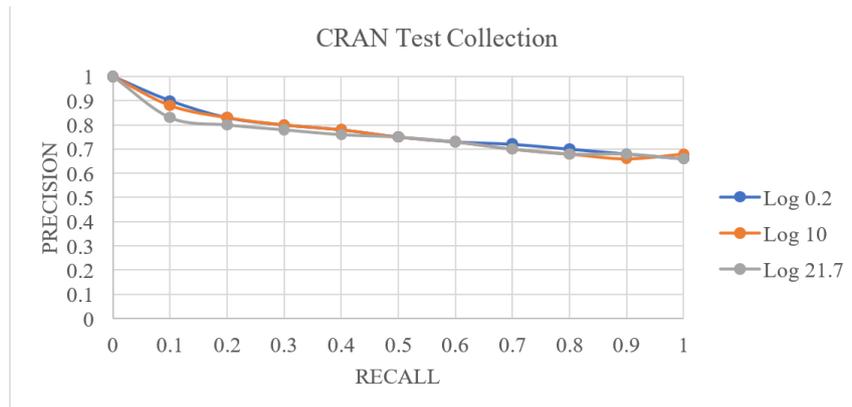

Figure 4. CRAN Best, Worst, and Standard Precision Values

It was determined the logarithm's base 0.2 is better than the standard log base 10. The assessment results in MAP@30 is equal 0.8825 at the best logarithm 0.2. The results are better than the standard logarithm at base 10 where MAP@30 value is equal to 0.8775. It was identified the worst performance at log base 21.7 for MAP@30 with a score of 0.8525. The best MAP value is equal to 0.7809 when we apply the TFIDF weighting technique using logarithm at base 0.3.

### 4.3. NPL Test Collection

The NPL test collection is the largest in the group of small test collections used for this research, it was found some improvements using log base 32.6, better precision values found at recall levels 0.2, 0.3, 0.4, 0.5, 0.6, 0.8, 0.9 and 1.0 in comparison with the standard log base 10, which is a very good improvement based on the data and in comparison, to other test collection results.

Table 10. NPL top results for MAP

| LOG | \multicolumn{11}{c}{RECALL} | MAP |
|---|---|---|---|---|---|---|---|---|---|---|---|---|
|  | 0 | 0.1 | 0.2 | 0.3 | 0.4 | 0.5 | 0.6 | 0.7 | 0.8 | 0.9 | 1 |  |
| 32.6 | 1 | 0.9 | 0.88 | 0.87 | 0.85 | 0.84 | 0.83 | 0.83 | 0.8 | 0.78 | 0.77 | 0.85 |
| 26.3 | 1 | 0.9 | 0.87 | 0.85 | 0.84 | 0.83 | 0.82 | 0.83 | 0.8 | 0.78 | 0.78 | 0.845455 |
| 55.4 | 1 | 0.9 | 0.87 | 0.85 | 0.84 | 0.83 | 0.82 | 0.83 | 0.8 | 0.78 | 0.78 | 0.845455 |
| 0.1 | 1 | 0.9 | 0.87 | 0.85 | 0.84 | 0.83 | 0.82 | 0.83 | 0.8 | 0.78 | 0.77 | 0.844545 |
| 1.8 | 1 | 0.9 | 0.87 | 0.85 | 0.84 | 0.83 | 0.82 | 0.83 | 0.8 | 0.78 | 0.77 | 0.844545 |

Above table defines log base 32.6 with MAP computed result of 0.85.

Table 11. NPL top results for MAP@30

| LOG | \multicolumn{11}{c}{RECALL} | MAP@30 |
|---|---|---|---|---|---|---|---|---|---|---|---|---|
|  | 0 | 0.1 | 0.2 | 0.3 | 0.4 | 0.5 | 0.6 | 0.7 | 0.8 | 0.9 | 1 |  |
| 32.6 | 1 | 0.9 | 0.88 | 0.87 | 0.85 | 0.84 | 0.83 | 0.83 | 0.8 | 0.78 | 0.77 | 0.9125 |
| 0.7 | 1 | 0.92 | 0.87 | 0.85 | 0.84 | 0.83 | 0.8 | 0.83 | 0.77 | 0.76 | 0.75 | 0.91 |
| 0.1 | 1 | 0.9 | 0.87 | 0.85 | 0.84 | 0.83 | 0.82 | 0.83 | 0.8 | 0.78 | 0.77 | 0.905 |
| 1.8 | 1 | 0.9 | 0.87 | 0.85 | 0.84 | 0.83 | 0.82 | 0.83 | 0.8 | 0.78 | 0.77 | 0.905 |
| 24.7 | 1 | 0.9 | 0.87 | 0.85 | 0.84 | 0.83 | 0.82 | 0.83 | 0.8 | 0.78 | 0.77 | 0.905 |

Also for MAP @30, it was found that the best result was gained when using the log base 32.6 with a score 0.9125.

Table 12. NPL comparison of MAP@30 of the best, worst and standard base 10

| LOG | \multicolumn{11}{c}{RECALL} | MAP@30 |
|---|---|---|---|---|---|---|---|---|---|---|---|---|
|  | 0 | 0.1 | 0.2 | 0.3 | 0.4 | 0.5 | 0.6 | 0.7 | 0.8 | 0.9 | 1 |  |
| 32.6 | 1 | 0.9 | 0.88 | 0.87 | 0.85 | 0.84 | 0.83 | 0.83 | 0.8 | 0.78 | 0.77 | 0.9125 |
| 10 | 1 | 0.9 | 0.87 | 0.85 | 0.84 | 0.83 | 0.8 | 0.83 | 0.78 | 0.77 | 0.76 | 0.905 |
| 26.3 | 1 | 0.9 | 0.87 | 0.85 | 0.84 | 0.83 | 0.83 | 0.82 | 0.8 | 0.78 | 0.78 | 0.905 |

The differences between the best, worst and the standard log base 10 results of MAP@30, are plotted in below graph in the figure to provide a better view of the results.

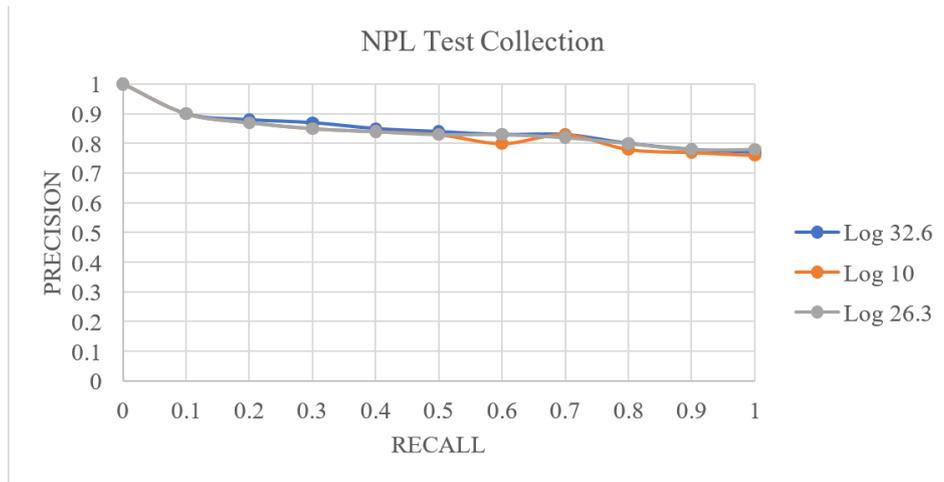

Figure 5. NPL test collection, best log at 32.6

The log base 32.6 has better precision at many recall levels, a second log base value of 0.7 was also determined having better MAP@30 compared to the standard log base 10.

### 4.4. LISA Test Collection

In this test collection, it was found other log bases that performs better than the standard base 10. Log base 49 has a score of 0.44 for MAP and 0.6925 for MAP@30 and is considered the best found per below data.

Table 13. LISA best log values for MAP

| LOG | \multicolumn{11}{c}{RECALL} | MAP |
|---|---|---|---|---|---|---|---|---|---|---|---|---|
| | 0 | 0.1 | 0.2 | 0.3 | 0.4 | 0.5 | 0.6 | 0.7 | 0.8 | 0.9 | 1 | |
| 49 | 1 | 0.72 | 0.55 | 0.5 | 0.48 | 0.42 | 0.42 | 0.29 | 0.18 | 0.15 | 0.13 | 0.44 |
| 49.1 | 1 | 0.72 | 0.55 | 0.5 | 0.48 | 0.42 | 0.42 | 0.29 | 0.18 | 0.15 | 0.13 | 0.44 |
| 85.1 | 1 | 0.72 | 0.55 | 0.5 | 0.48 | 0.42 | 0.42 | 0.29 | 0.18 | 0.15 | 0.13 | 0.44 |
| 0.3 | 1 | 0.72 | 0.55 | 0.5 | 0.48 | 0.42 | 0.42 | 0.29 | 0.18 | 0.14 | 0.13 | 0.439091 |
| 0.4 | 1 | 0.72 | 0.55 | 0.5 | 0.48 | 0.42 | 0.42 | 0.29 | 0.18 | 0.14 | 0.13 | 0.439091 |

Table 14. LISA best log values for MAP@30

| LOG | \multicolumn{11}{c}{RECALL} | MAP@30 |
|---|---|---|---|---|---|---|---|---|---|---|---|---|
| | 0 | 0.1 | 0.2 | 0.3 | 0.4 | 0.5 | 0.6 | 0.7 | 0.8 | 0.9 | 1 | |
| 49 | 1 | 0.72 | 0.55 | 0.5 | 0.48 | 0.42 | 0.42 | 0.29 | 0.18 | 0.15 | 0.13 | 0.6925 |
| 49.1 | 1 | 0.72 | 0.55 | 0.5 | 0.48 | 0.42 | 0.42 | 0.29 | 0.18 | 0.15 | 0.13 | 0.6925 |
| 85.1 | 1 | 0.72 | 0.55 | 0.5 | 0.48 | 0.42 | 0.42 | 0.29 | 0.18 | 0.15 | 0.13 | 0.6925 |
| 0.3 | 1 | 0.72 | 0.55 | 0.5 | 0.48 | 0.42 | 0.42 | 0.29 | 0.18 | 0.14 | 0.13 | 0.6925 |
| 0.4 | 1 | 0.72 | 0.55 | 0.5 | 0.48 | 0.42 | 0.42 | 0.29 | 0.18 | 0.14 | 0.13 | 0.6925 |

The table above indicates top log values for MAP and MAP@30, it was found that the best MAP is at log 49 with a MAP score 0.85.

Table 15. LISA best and worst values compared to the standard base 10

| LOG | RECALL | | | | | | | | | | | MAP@30 |
|---|---|---|---|---|---|---|---|---|---|---|---|---|
| | 0 | 0.1 | 0.2 | 0.3 | 0.4 | 0.5 | 0.6 | 0.7 | 0.8 | 0.9 | 1 | |
| 49 | 1 | 0.72 | 0.55 | 0.5 | 0.48 | 0.42 | 0.42 | 0.29 | 0.18 | 0.15 | 0.13 | 0.6925 |
| 10 | 1 | 0.72 | 0.51 | 0.5 | 0.45 | 0.42 | 0.42 | 0.29 | 0.18 | 0.13 | 0.11 | 0.6825 |
| 0.6 | 1 | 0.72 | 0.51 | 0.5 | 0.45 | 0.42 | 0.42 | 0.29 | 0.18 | 0.15 | 0.14 | 0.6825 |

The above table's data is shown as a graph in the below figure.

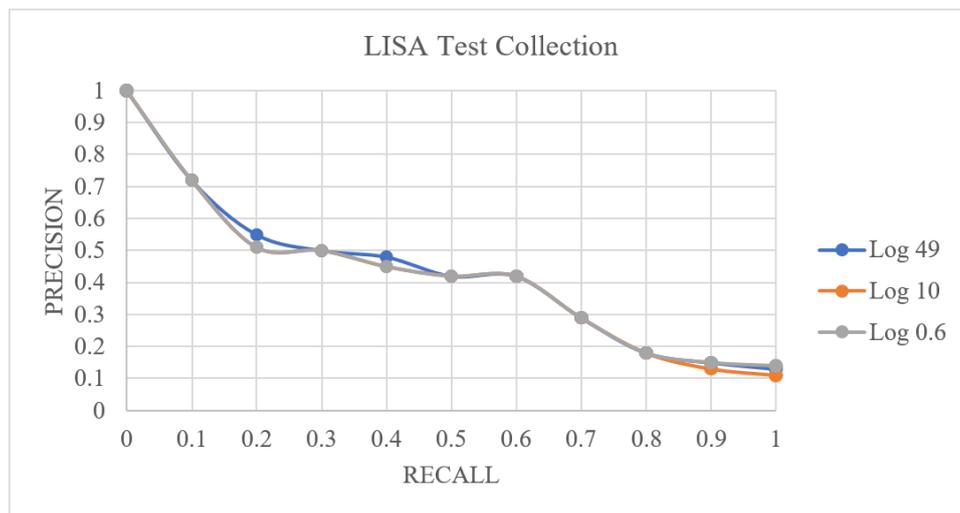

Figure 6. List best and worst compared with log base 10

At recall levels 0.2, and 0.4, for example, the curve is higher than the other log bases compared. Therefore, log base 49 is considered better than the others. This has improvements in the performance for MAP@30 value equal to 0.6925. The results are better than the standard logarithm at base ten where the MAP@30 value is equal to 0.6825. We found the worst performance was of log base 0.6 and it has the same MAP@30 value of 0.6825 compared to the standard log base 10.

### 4.5. CISI Test Collection

It was also used to find better results using different log bases for this test collection. Log 84.6 have a MAP@30 score of 0.87 and MAP score of 0.7136. Its precision is higher at recall levels 0.1, 0.2, 0.5, 0.7 and 0.8.

Table 16. CISI best 5 log values MAP

| LOG | RECALL | | | | | | | | | | | MAP |
|---|---|---|---|---|---|---|---|---|---|---|---|---|
| | 0 | 0.1 | 0.2 | 0.3 | 0.4 | 0.5 | 0.6 | 0.7 | 0.8 | 0.9 | 1 | |
| 84.6 | 1 | 0.9 | 0.83 | 0.75 | 0.72 | 0.68 | 0.63 | 0.62 | 0.61 | 0.56 | 0.55 | 0.713636 |
| 0.7 | 1 | 0.88 | 0.75 | 0.73 | 0.72 | 0.67 | 0.66 | 0.63 | 0.61 | 0.6 | 0.6 | 0.713636 |
| 53.1 | 1 | 0.88 | 0.75 | 0.73 | 0.72 | 0.67 | 0.66 | 0.63 | 0.61 | 0.6 | 0.6 | 0.713636 |
| 58.2 | 1 | 0.88 | 0.75 | 0.73 | 0.72 | 0.67 | 0.66 | 0.63 | 0.61 | 0.6 | 0.6 | 0.713636 |
| 58.3 | 1 | 0.88 | 0.75 | 0.73 | 0.72 | 0.67 | 0.66 | 0.63 | 0.61 | 0.6 | 0.6 | 0.713636 |

Table 17. CISI best 5 log values MAP @30

| LOG | Recall | | | | | | | | | | | MAP@30 |
|---|---|---|---|---|---|---|---|---|---|---|---|---|
| | 0 | 0.1 | 0.2 | 0.3 | 0.4 | 0.5 | 0.6 | 0.7 | 0.8 | 0.9 | 1 | |
| 84.6 | 1 | 0.9 | 0.83 | 0.75 | 0.72 | 0.68 | 0.63 | 0.62 | 0.61 | 0.56 | 0.55 | 0.87 |
| 91.8 | 1 | 0.88 | 0.8 | 0.75 | 0.72 | 0.66 | 0.63 | 0.61 | 0.6 | 0.6 | 0.58 | 0.8575 |
| 94 | 1 | 0.88 | 0.8 | 0.75 | 0.72 | 0.66 | 0.63 | 0.61 | 0.6 | 0.6 | 0.58 | 0.8575 |
| 0.9 | 1 | 0.88 | 0.8 | 0.75 | 0.72 | 0.66 | 0.63 | 0.61 | 0.6 | 0.6 | 0.55 | 0.8575 |
| 58 | 1 | 0.88 | 0.8 | 0.75 | 0.72 | 0.66 | 0.63 | 0.61 | 0.6 | 0.6 | 0.55 | 0.8575 |

When we assess the CISI test collection, we find that the system is at its best performance with a MAP@30 value equal to 0.87 when we apply the TFIDF weighting technique using logarithm base equal to 84.6. The results are better than the standard logarithm at base ten where the MAP@30 value is equal to 0.8575.

Table 18. CISI Comparison of Best and Worst with the Standard log 10.

| LOG | RECALL | | | | | | | | | | | MAP@30 |
|---|---|---|---|---|---|---|---|---|---|---|---|---|
| | 0 | 0.1 | 0.2 | 0.3 | 0.4 | 0.5 | 0.6 | 0.7 | 0.8 | 0.9 | 1 | |
| 84.6 | 1 | 0.9 | 0.83 | 0.75 | 0.72 | 0.68 | 0.63 | 0.62 | 0.61 | 0.56 | 0.55 | 0.87 |
| 10 | 1 | 0.88 | 0.8 | 0.75 | 0.72 | 0.66 | 0.63 | 0.61 | 0.6 | 0.6 | 0.55 | 0.8575 |
| 85.1 | 1 | 0.9 | 0.72 | 0.66 | 0.63 | 0.61 | 0.6 | 0.6 | 0.55 | 0.54 | 0.52 | 0.82 |

Based on the above table, improvements has been determined by the results of calculation for MAP@30 and it is illustrated in below figure the graph.

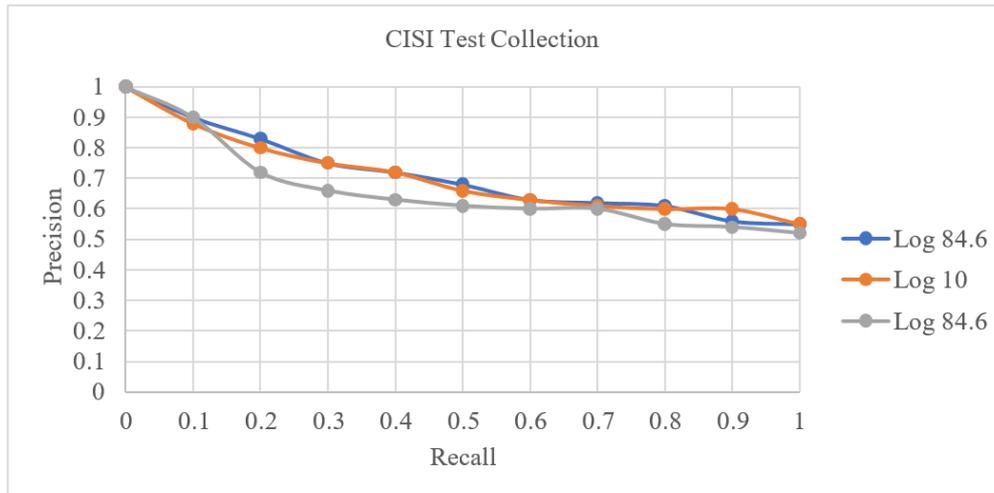

Figure 7. CISI Best, Worst, and Standard based comparison on MAP@30

## 5. CONCLUSIONS

The experiments show that calculating logarithm's using base 10 is not always the optimal solution and it was found new values to calculate the TFIDF in all five test collections, resulting in improvements in the matter of efficiency and effectiveness of the information retrieval system. It has been proved in this research that a variation in the logarithm's bases improves the precision at high, middle and low recall levels, which is considered a good improvement. Many log bases outperforms log base 10. Log values such as 84.6, 49, 32.6, 0.2 and 0.1 had an impact for better performance in the system. It must be taken in consideration that in the test collections CISI, LISA and NPL, better results were found using log base higher than 10 and that MED and CRAN had better results with log bases lower than base ten. I conclude that in this research, a variant of the logarithms bases in the TFIDF weighting technique affects the performance of the results and it was found in the context. It has been determined which log base returns better results for each of the test collections that was used. Further research must be done on large test collections such as TREC, GOV. The method of changing the log base should also be applied in a probabilistic model.

### ACKNOWLEDGEMENTS

I would like to thank everyone that have taught me a lesson in my life, specially authors of scientific papers and books.

## AUTHOR

**Kamel Assaf** is a programmer in the I.T. department for a company in Edmonton, Ab, Canada. He has completed his M.Sc. in Computer Science from the Lebanese International University. His research interest lies in the area of Information Retrieval, Data Mining and Computer Graphics

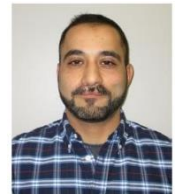